# Silicon Photonics Mode-Selective Phase Shifter

Seyed Mohammad Reza Safaee, *Graduate Student Member, IEEE*, Kaveh (Hassan) Rahbardar Mojaver, *Member, IEEE,* and Guowu Zhang, *Graduate Student Member,* Odile Liboiron-Ladouceur, *Senior Member, IEEE*

*Abstract*— A novel mode-selective thermo-optic phase shifter (MS-TOPS) enabled by subwavelength grating (SWG) structures is proposed and experimentally demonstrated on a 220 nm waveguide thick silicon photonics chip for the first two quasi-transverse electric modes (TE0, TE1). Mode-selective relative phase manipulation of modes unlocks several processing tasks in mode division multiplexing systems. This integrated solution provides a direct phase manipulation of modes without converting them to their fundamental modes. A Mach-Zehnder interferometer is deployed as a test structure incorporating the proposed MS-TOPS in one arm and a mode-insensitive thermo-optic phase shifter (MI-TOPS) in another. The effect of the SWG duty cycle ratio is investigated by both numerical simulations and experimental measurements. A mode-selectivity of 1.44 is experimentally demonstrated. In other words, the thermo-optic coefficient of TE0 is 44% larger than the one for TE1. The phase shifter's insertion loss is at most 2.5 dB and a worst-case crosstalk of -13.1 dB over a 40 nm wavelength range from 1520 to 1560 nm. A cascaded configuration of the proposed MS-TOPS and an MI-TOPS provides sufficient degrees of freedom to manipulate the relative phase of each mode independently. Potential numerous applications of such devices include optical switching, multimode quantum optical processors, and scaling-up conventional optical processors with a mode-selective building block.

## I. INTRODUCTION

Extensive growth of silicon photonics in the past decade is not only owing to its compatibility with the widespread CMOS (complementary metal-oxide-semiconductor) fabrication technology but also on account of its potential for enabling high-speed data transfer, reduced power consumption, and improved performance compared to its counterpart solutions [1]. Deployment of the silicon-on-insulator (SOI) fabrication process provides a relatively large refractive index difference between the core (silicon, Si), and cladding (silicon dioxide, $SiO_2$) materials, which can confine different orthogonal eigenmodes inside the SOI waveguide. Historically, the dominant approach in silicon photonics devices was to operate in single-mode, below the cut-off condition of higher-order modes, due to significant levels of inter-modal crosstalk, caused by the underdeveloped components required for multimode operation [2]. Lately, the advantageous performance of the mode division multiplexing (MDM) scheme in terms of further expanding the optical link capacity in an energy efficient approach has motivated significant research towards developing on-chip MDM components [3]-[5].

Conventional on-chip MDM architectures do not predominantly provide individual access to the higher-order modes that are intended to be processed (e.g., switching, modulation). Instead, the prevailing approach typically involves encoding of higher modes into their fundamental mode, followed by the execution of the desired processing function using single-mode components, and reconverting them to their primary higher-order mode profile [6], [7]. Although this approach provides a straightforward solution for developing numerous MDM applications, it requires a bulky (de)multiplexing photonic circuitry that inevitably deteriorates insertion loss, power consumption, and crosstalk. In contrast, state-of-the-art MDM research involves efforts to provide building blocks that enable direct manipulation of modes such as the proposed approaches for space-based and intra-mode switching applications [8]-[11].

In contrast to the limited applicability of the previous research efforts, individual phase manipulation of each mode is a highly versatile optical functionality in MDM systems. This is the cornerstone of realizing almost every desired processing task from switching, and filtering to photonic-based vector/matrix-based mathematical multiplication. For example, by directly controlling the relative phase between different modes without encoding them to their fundamental mode distribution, the need for mode (de)multiplexing is eliminated, which can support a more efficient design in terms of footprint, crosstalk, and potentially power consumption. In this work, direct phase manipulation of modes is demonstrated for the first time. As a proof of concept, the phase of the first and second order quasi transverse electric (TE) mode profiles (TE0, TE1) are manipulated by the proposed mode-selective thermo-optic phase shifter (MS-TOPS). Mode-selectivity as a design figure-of-merit (FoM) is realized by deploying subwavelength grating

Manuscript submitted for review on July 28, 2023.
This work was supported by the Natural Sciences and Engineering Research Council of Canada (NSERC).
S. M. R. Safaee, K. Rahbardar Mojaver, and O. Liboiron-Ladouceur are with the Department of Electrical and Computer Engineering, McGill University, Montreal, QC, H3A 0E9, Canada. (emails: seyed.safaeeardestani@mail.mcgill.ca, hassan.rahbardarmojaver@mcgill.ca, odile.liboiron-ladouceur@mcgill.ca)
G. Zhang was with McGill University, Montreal, QC, Canada. He is now with Intelligent Optics & Photonics Research Center, Jiaxing Institute of Zhejiang University, Jiaxing 314031, China. (e-mail: guowu.zhang@zju.edu.cn).

(SWG) structures that provide a method to engineer the waveguide refractive index for different modes. We verify the effect of the SWG geometric parameters both analytically and using two-dimensional (2D) numerical simulations. We experimentally validated the working principle of the MS-TOPS by investigating the effect of changing the SWG duty cycle ratio. We introduce an MZI test structure incorporating the MS-TOPS in one arm, and a mode-insensitive TOPS in another as an additional degree of freedom to manipulate both modes independently. We provide a methodology to experimentally extract the designed FoM and compare it with the three-dimensional (3D) simulation results. Finally, we briefly review the potential applications of the proposed MS-TOPS as a highly versatile and enabling building block in MDM systems.

## II. DESIGN THEORY AND ANALYSIS

### A. Thermo-Optic Phase Shifter (TOPS)

Integrated tunable phase shifters in the SOI platform operate based on altering the effective refractive index. This is commonly realized by making use of thermo-optic effect, electro-optic effect (also known as free-carrier plasma dispersion effect), or microelectromechanical system (MEMS) actuation. Despite the lower power consumption of the latter approach, MEMS fabrication is non-standard in today's silicon photonics technology. In contrast to TOPS, electro-optic phase shifters offer a notably larger modulation bandwidth at the expense of higher insertion loss, which separates the target applications of each methodology. In particular, TOPS plays a significant role in various reconfigurable silicon photonics applications [12] such as optical neural networks [13], and mode-division multiplexing systems [14].

TOPS working principle relies on phase of light alteration that is induced by a heater-imposed temperature change of $\Delta T$ (i.e., thermo-optic effect). For a heated waveguide with length $L$, the temperature-dependent phase change at $\lambda_0$ free-space wavelength is [15]:

$$\Delta \Phi = \frac{2\pi L}{\lambda_0} \frac{dn_{eff}}{dT} \Delta T \quad (1)$$

where $dn_{eff}/dT$ is the thermo-optic coefficient of the waveguide. Thermo-optic coefficient of silicon and silicon dioxide are respectively approximately $1.86 \times 10^{-4} \text{K}^{-1}$ and $0.95 \times 10^{-5} \text{K}^{-1}$ at T = 300 K, and $\lambda_0 = 1550$ nm [16], which shows more than an order of magnitude difference between waveguide core and cladding material. This difference implies that the waveguide geometries with respect to the guided mode profile can have an impact on the thermo-optic coefficient of the waveguide. This phenomenon provides an opportunity to engineer the thermo-optic coefficient of the waveguide in order to differ for various modes as an endeavor to reach mode-selective TOPS.

We previously investigated the TOPS waveguide width effect on altering the thermo-optic coefficient for the two fundamental TE modes using both numerical simulation tools and an experimental validation methodology [17]. Considering a 220 nm thick, and 0.96 μm wide waveguide, the design yields about nine percent larger $dn_{eff}/dT$ for the TE1 compared to the TE0 mode (1.96 versus 1.80, respectively) demonstrating a proof-of-concept MS-TOPS although not sufficiently efficient for practical applications. On the other hand, a waveguide width of 4 μm provides only a difference of less than 1% in terms of thermo-optic coefficient for the two fundamental TE modes, which creates a desirable waveguide width to build mode-insensitive TOPS (MI-TOPS) [18].

As a design FoM, mode-selectivity is defined as the ratio of thermo-optic coefficients of the first and second fundamental TE modes (TE0, and TE1):

$$\zeta = \frac{dn_{eff}(TE0)/dT}{dn_{eff}(TE1)/dT} = \frac{P_{\pi,TE1}}{P_{\pi,TE0}} \quad (2)$$

where $P_{\pi,TE1}$, and $P_{\pi,TE0}$ are the required heating power to produce an electro-optic phase shift of 180° (π) for the respective modes. The latter fraction in the above equation is derived considering a linear relation between the heating power

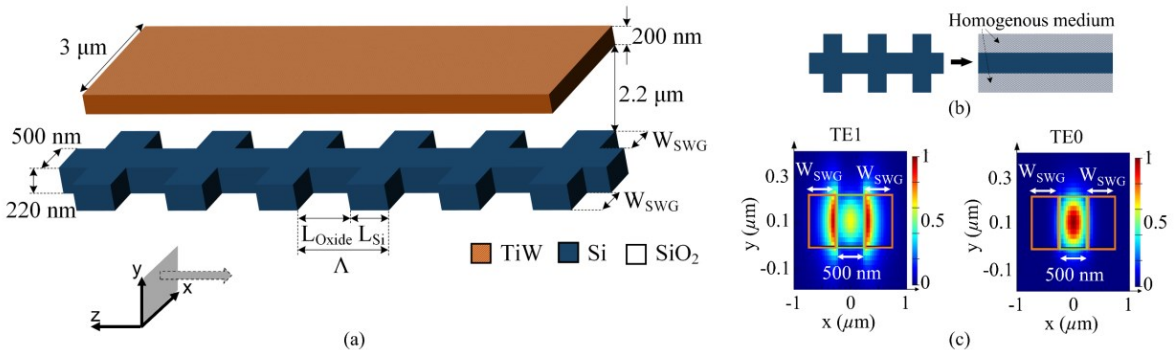

Figure 1 (a) The schematic diagram of the proposed SWG-based TOPS; (b) Modeling the corrugated sections as a homogenous medium with an engineered refractive index represented by Eq. 3 for the 2D simulation objective; (c) Simulated (2D) normalized electric field distribution for both TE0, and TE1 modes in the xy plane (perpendicular to the propagation axis).

and temperature change in Eq. (1), *i.e.*, $P_\pi \propto (dn_{eff}/dT)^{-1}$. This relation suggests a straightforward method for the MS-TOPS mode-selectivity extraction using solely a transmission measurement from an interferometric structure while sweeping the applied heating power.

*B. Subwavelength grating-based TOPS*

Silicon subwavelength grating structures (also known as SWG metamaterials) have extensive range of applications in integrated photonic devices including fiber-chip surface and edge grating couplers [19], [20], broadband contra directional couplers [21], mode-selective add-drop couplers [22], and several others [23]. SWGs enable engineering the refractive index of strip waveguides by behaving as a homogenous medium with a range of possible effective refractive index [24]. This property of the SWGs is exploited to design a mode-selective TOPS as demonstrated in Fig. 1(a). In this work, a SWG structure with a periodicity (pitch) of $\Lambda$ is considered on both sides of the conventional single-mode strip waveguide (500 nm wide, and 220 nm thick). The SWG periodicity is well below the Bragg condition remaining in the sub-wavelength regime [24]. The duty cycle ratio of the grating, $f = L_{Si}/\Lambda$, is the volume fraction of silicon where $L_{Si}$ is the length of the un-etched silicon in the SWG structure. Based on the effective medium theory, the side corrugation structures could be considered as a homogenous medium as shown in Fig. 1(b). Considering a TE polarization beam with almost no electric field component in the propagation direction, the engineered refractive index of the corrugated structure is well approximated as [24]:

$$n_{SWG} = \sqrt{f \cdot n_{Si}^2 + (1-f) \cdot n_{Oxide}^2} \ , \quad (3)$$

where $n_{Si} = 3.46$ and $n_{Oxide} = 1.45$ are the silicon and silicon dioxide refractive index at $\lambda_0 = 1550$ nm, respectively. For an arbitrary duty ratio *f* of 0.4, the SWG refractive index reaches a value of 2.46, which implies a potentially lower thermo-optic coefficient for the SWG compared to the waveguide core.

As supported by the electric field modal distribution results, attained by Ansys Lumerical photonic simulation tool, provided in Fig. 1 (c), the TE0 main lobe propagates through the un-corrugated silicon waveguide core while TE1 mode lobes mostly propagate through the sides of the waveguide where the SWG structure exists. Considering the propagation behavior of the TE1 mode, when heat applied through the deployed titanium-tungsten alloy (TiW) metal heater shown in Fig. 1 (a), this lower thermo-optic coefficient of the SWG for TE1 translates eventually to a smaller imposed phase shift on the TE1 mode in contrast with the TE0 mode enabling a mode-selective TOPS.

As a proof-of-concept design, the SWG is 450 nm wide ($W_{SWG}$ = 450 nm). The metal heater is deposited 2.2 μm above the waveguide extending 3 μm wide to ensure a uniform heating. The SWG width should be optimized for to the application requirements. Minimizing the SWG width improves the heater energy efficiency at the cost of a lower mode-selectivity and an increased TE1 mode insertion loss due to a relative increase in scattering from sidewall roughness. On the other hand, an increased SWG width can provide a greater TE1 mode confinement in the SWG region rather than the waveguide core area leading to a larger $dn_{eff}/dT$ contrast between two modes. This could also lower the sidewall roughness scattering loss contribution in the TE1 mode insertion loss.

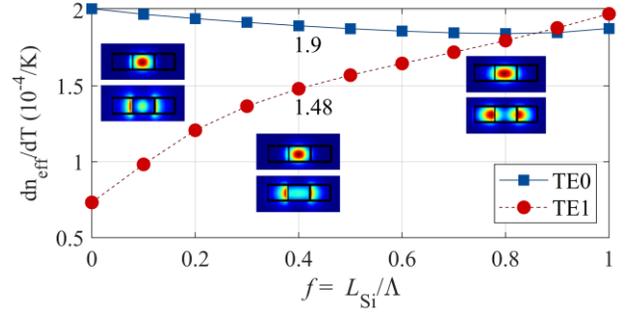

Fig. 2. Thermo-optic coefficient of the proposed mode-sensitive TOPS for TE0 and TE1 mode profiles versus duty cycle ratio calculated using a 2D simulation approach ($W_{SWG}$=450 nm, $\Lambda$=220 nm). Subsets show the electric field intensity of a cross section perpendicular to the propagation direction for *f*= {0.1, 0.4, 0.8}.

Minimizing the SWG duty ratio is favored because a greater *f* reduces the TOPS $dn_{eff}/dT$ contrast between two modes; hence, reaching the maximum mode-selectivity (*i.e.*, lowest *f*) is limited by the minimum feature size of the fabrication technology. Increasing the SWG pitch facilitates attaining a smaller duty ratio well above the minimum feature size. However, staying in the subwavelength regime and avoiding the Bragg reflection condition determines the upper boundary limit of the pitch ($\Lambda < \lambda_{min}/2n_{eff}$). Considering $\Lambda$ = 220 nm as a proof-of-concept design, the impact of the duty ratio is investigated by conducting a 2D simulation approach using Ansys Lumerical finite difference eigenmode solver tool. In this approach, Eq. 3 represents the corrugated sections as an effective medium besides the silicon waveguide core, while the whole structure is elongated toward infinity in the propagation direction. The effective refractive index of the structure is attained in room temperature, and following a 100K temperature increase. Assuming $dn_{eff}/dT \simeq \Delta n_{eff}/\Delta T$ in that temperature range, the thermo-optic coefficient is calculated for each mode. Results for different duty ratios are depicted in Fig. 2 supporting a mode-selectivity of $\zeta$=1.28 for *f* =0.4, corresponding to $L_{Si} = 88$ nm being marginally above the minimum feature size of the accessible technology.

Upon optimization of different dimensions using 2D simulation approach, a 3D finite difference time domain (FDTD) simulation is carried out using Ansys Lumerical FDTD solver tool. This involves calculation of effective refractive indices from extracting the phase velocity, which requires attaining the dispersion diagram of the SWG TOPS (also called a band structure analysis) [25]. More specifically, Fourier transform of some randomly placed time-domain monitors in the simulation region leads to the frequency-wavenumber relation as the mode guiding condition (*i.e.*, the dispersion diagram) providing the phase velocity parameter. The calculated mode-selectivity FoM for $f$ = {0.4, 0.5, 0.6} is compared with experimental results in section III. B.

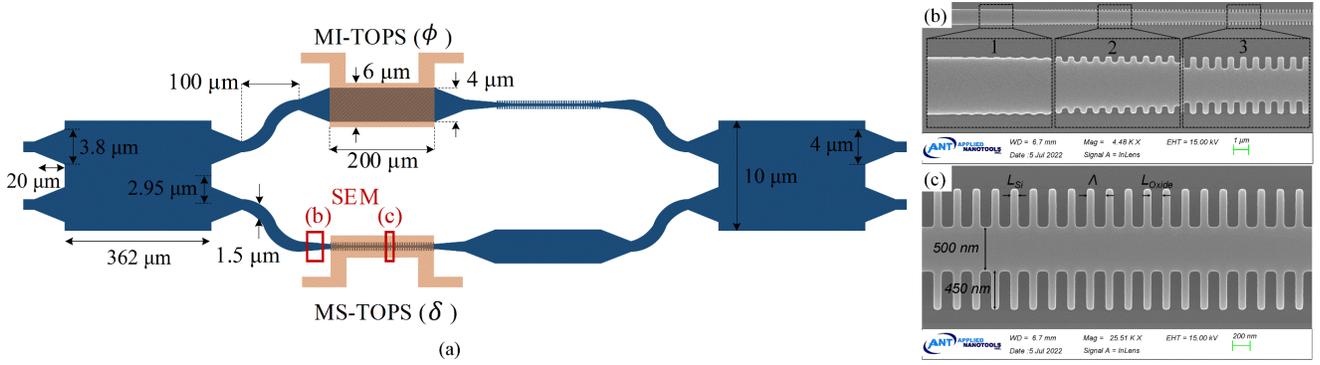

Fig. 3: Schematic diagram of the balanced interferometric test structure designed for the evaluation of the SWG-based MS-TOPS ($\delta$ phase shifter); (b) SEM image of a typical designed taper that provides a low loss interconnection between a 250 nm SWG waveguide and neighboring components (subsets 1-3 are enlarged for better visibility); (c) SEM image of the designed SWG waveguide with dimensions: $W_{SWG}$ = 450 nm, $\Lambda$ = 220 nm, $f$ = 0.4, $L_{Si}$ = 88 nm.

*C. Interferometric test structure*

The conventional TOPS test structure involves using a balanced Mach-Zehnder interferometer (MZI), which incorporates the same TOPS design in both arm with one typically operational and the other making the MZI loss-balanced. Tuning the TOPS alters the light phase in one arm and modulates the output transmission intensity in an interferometric manner. The schematic diagram of the MS-TOPS test structure is demonstrated in Fig. 3(a). The 2×2 MZI structure is comprised of two multimode interferometers (MMIs), 100 μm long low-loss S-bend at its output to thermally set the MZI arms apart, and a mode-insensitive thermo-optic phase shifter (MI-TOPS), and tapers to provide a low-loss interconnection between components with different waveguide widths. Excluding the SWG structure, and the tapers at both ends of the SWG, all geometric parameters are optimized based on a previous work [18] as illustrated in Fig. 3 (a).

Fig. 3 (b) shows the SEM image of the fabricated taper used at both ends of a 250 nm wide SWG waveguide (the optimized design is 450 nm wide). This linear taper from the single-mode waveguide to the SWG waveguide was optimized using the design methodology reported in [26]. The taper length was swept using a 3D FDTD method to ensure a smooth transition between the conventional waveguides and the SWG structure realizing above 99.6% power transmission for both TE0 and TE1 modes with a 20 μm long taper [26].

The 200 μm long MI-TOPS waveguide is 4 μm wide providing similar thermo-optic coefficient for TE0 and TE1 modes, within a contrast of 2% ($\pm 1$%). Deploying such a component enables a more comprehensive evaluation of the MS-TOPS performance by imposing an adjustable $\phi$ phase shift on both modes in one arm while applying another adjustable $\zeta \cdot \delta$ and $\delta$ phase shift on TE0 and TE1 modes, respectively.

Fig. 3(c) shows the SEM image of the fabricated SWG-based MS-TOPS validated using the interferometric structure. The fabrication in an SOI platform is performed using Applied Nanotools (ANT) NanoSOI fabrication process. The optical waveguides are 220 nm thick fully etched on top of 2 μm buried oxide (BOX). The high resistive titanium-tungsten (TiW) alloy functions as metal heaters deposited 2.2 μm on top of waveguides. A low resistive titanium-tungsten/aluminum bi-layer (TiW/Al) enables the contact with heaters and metal routings.

### III. EXPERIMENTAL DEMONSTRATION

*A. Validation Testbed*

The fabricated MS-TOPS in the interferometric test structure is the device under test (DUT) in the relatively straightforward testbed, schematically illustrated in Fig. 4. The DUT is optically probed by a multi-channel single mode fiber array unit (FAU) using on-chip vertical grating couplers from ANT process design kit (PDK). The bare die is electrically probed with a multi-contact wedge (MCW). Because the repeatability of the measurement depends on the MCW contact resistance varying with each probe landing attempt, the power supply in the current mode eliminates the dependency of the applied bias and phase shift relation to the contact resistance within the range of current source stability. An on-chip electrical loopback between two electrical pads facilitates secure and more repeatable landing and connection to the die of the MCW.

In the first measurement scenario, an external directional coupler (DC) evenly splits (*i.e.*, splitting ratio of $S_1$=50:50) the power of the tunable laser source to enable simultaneous TE0 and TE1 mode propagation through the DUT. The second order mode is excited on-chip using adiabatic coupler-based mode multiplexers [27]. Embedded $\phi$ (mode-insensitive), and $\delta$ (mode-selective) TOPS in each arm of the MZI DUT are biased and swept according to their metal heater resistivity to deliver desired heat. Sweeping the applied current to each phase shifter creates a two-dimensional space to measure the DUT, which further enables calculating its mode-selectivity as the main design FoM. In the second measurement scenario, an external DC with an $S_2$=10:90 splitting ratio is deployed. The objective of this scenario is to extract the DUT crosstalk and insertion loss for each mode; hence, it requires two consecutive measurements: once for TE0 while TE1 input is null and vice versa. In each case, the output is de-multiplexed on-chip to the primary and crosstalk signals, which are simultaneously measured with two optical power meters. The 10% splitting branch is fed to the on-chip alignment loopbacks promoting a more robust and accurate insertion loss assessment. The insertion loss and modal crosstalk measurement accuracy is masked by the FAU random drift of $\pm 0.2$ dB during a ten-minutes time window of stability measurement. Moreover, within the same duration, a random drift of $\pm 0.3$ mW in the delivered electrical power to the heaters limits the phase shift controllability to 0.09 rad.

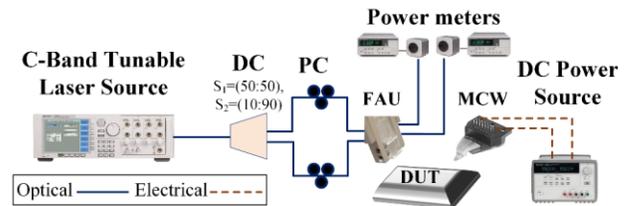

Fig. 4: Schematic block diagram of the testbed consisting of an external directional coupler (DC) in two different scenarios ($S_1$ for simultaneous, and $S_2$ for successive mode propagation through the DUT), polarization controller (PC), fiber array unit (FAU) optical probe, device under test (DUT), and multi-contact wedge (MCW) electrical probe.

### B. Measurement results and discussion

The DUT characterization objective is to measure the mode-selectivity of the designed MS-TOPS as defined in Eq. (2) being the principal design FoM. Using the discussed testbed (Fig. 4 with $S_1$=50:50), the phase shift in both arms of the MZI DUT is altered while the output bar port transmission is monitored. As elaborated in the 2D transmission contour maps in Fig. 5(a), sweeping the applied current ($I_\phi$) of the MI-TOPS (vertical axis) yields the same behavior for both TE0, and TE1 modes, as expected, for any constant $\delta$ phase shift. In other words, the required current change for inducing a $2\pi$ phase change (*e.g.*, distance between two bias points corresponding to minimum transmission) stays the same regardless of the mode (i.e., $P_{\pi,TE0} = P_{\pi,TE1}$). This confirms the mode insensitivity of the $\phi$ phase shifter. On the other hand, sweeping the MS-TOPS current ($I_\delta$ in the horizontal axis) demonstrates a different behavior for each mode profile (i.e., $P_{\pi,TE0} \neq P_{\pi,TE1}$). A dash-dotted line shows the applied current $I_\delta$ for the first two transmission deeps to provide a visual reference point. Distance between these two lines represents the additional required current for achieving a $2\pi$ phase shift. As shown in Fig. 5 (a), this distance in the TE1 results is greater than the TE0 outcome confirming the DUT mode-selectivity behavior.

Quantitative validation of the MS-TOPS behavior leads to extracting the mode-selectivity FoM. To accurately extract $P_\pi$, the normalized linear output power transmission of both modes is plotted versus the $\delta$ phase shifter heating power at a constant phase $\phi$ (*e.g.*, $I_\phi = 0$ as illustrated at the bottom of Fig. 5 (a) with down sampled data points for a better distinctness). This output optical power of the MZI DUT at the bar port conforms with the expected $\sin^2(\delta/2 + \delta_{OS})$ relation where $\delta_{OS}$ represents the observed phase offset at a zero-phase shift due to fabrication imperfection with the DUT second input port nulled. The fitted dashed curve facilitates the $P_\pi$ extraction as half of the distance between two consecutive deeps in the corresponding graph. Considering the experimental results of the DUT with a duty ratio of $f = 0.4$ as shown in Fig. 5 (a), this procedure produces a mean mode-selectivity of $\zeta = 1.44 \pm 0.05$ for a total number of 18 measured $\phi$ data points as $I_\phi = [0:2:36]\ mA$. The term $\pm 0.05$ shows the range of extracted FoM among these 18 measurements. The source of error in this mode-selectivity assessment is ascribed to the FAU and MCW random drift discussed in the testbed validation section (III.A). Moreover, the parameter extraction using the mathematical relation $\sin^2(\delta/2 + \delta_{OS})$ fits a large number of the measurement data leading to a more accurate FoM extraction reducing this uncertainty by an order of magnitude.

For a behavioral verification purpose, a similar 2D contour map is analytically calculated and illustrated in Fig. 5(b) for the same interferometric test structure ($\zeta = 1.44$) for both mode profiles under study. The MZI bar transmission follows the relation of $0.5 \times [\exp(j\phi) - \exp(j\delta)]$ times the input signal considering a null second input port. Considering mode insensitive phase $\phi$, and $\delta$ being the TE0 phase shift in each arm of the MZI, the relation becomes $0.5 \times [\exp(j\phi) - \exp(j\delta/\zeta)]$ for the TE1 mode at the bar output. Assuming a linear relationship between TOPS dissipated electrical power and the induced temperature gradient, the TOPS phase shift is proportional to the square of the heater current. As a result, the applied current and heating power units in Fig. 5(b) are arbitrary units (a.u.). This square relation considered between the applied current and the resulting phase shift is a realistic simplification of the heating power distribution. The resultant 2D contour map conforms with the experimental behavior. In comparison, it is worth noting that the maximum bar output transmission happens analytically at a $\pi$ phase shift. In other words, the bar output is null for a fully balanced MZI (at zero bias); however, experimental results show existence of a power offset at zero bias due to the phase offset originated from fabrication process variation that makes the MZI unbalanced. This could also be ascribed to the MMI performance deviation from the 50:50 splitting. Furthermore, TE0 and TE1 phase mismatch at the output of the on-chip coupler-based mode multiplexer contributes toward creating a different $\delta_{OS}$ for each mode. Depending on the target application, deploying a $1\times 2$ MMI with a balanced phase shift from the input to both outputs could promote the DUT power consumption efficiency.

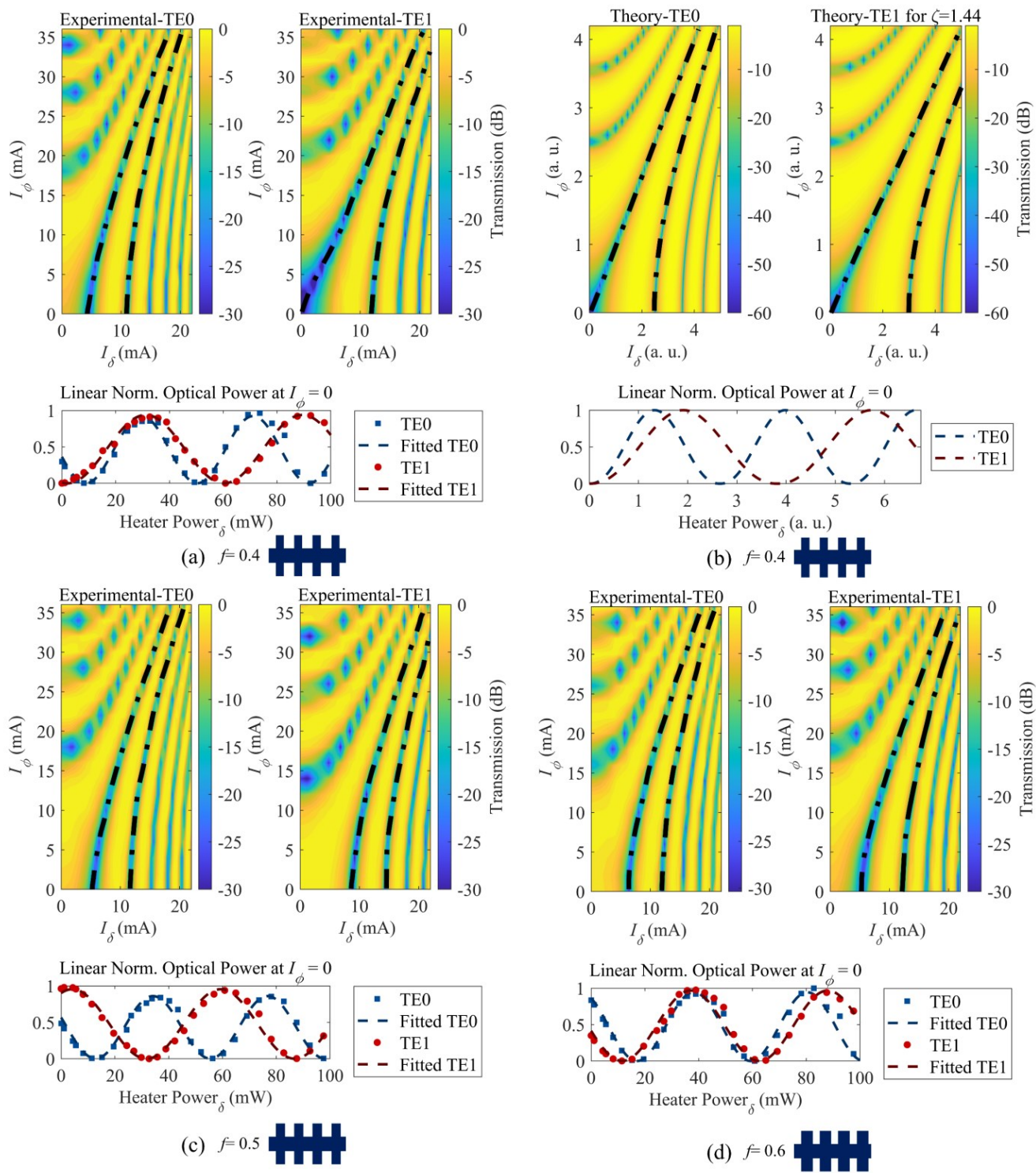

Fig. 3: The 2D contour map of the DUT bar output transmission versus applied current to the MS-TOPS (horizontal axis, $I_\delta$), and the MI-TOPS (vertical axis, $I_\phi$) for different SWG duty ratios. The dash-dotted line on each 2D contour map illustrates the first two transmission deeps as a visual reference point for comparing the DUT behavior in dealing with TE0 and TE1 modes. At the bottom of each 2D graph, the linear normalized optical power versus MS-TOPS heater power at $I_\phi = 0$ is shown where markers and dashed curves represent the down sampled experimental data points and the fit with $\sin^2(\delta/2 + \delta_{OS})$ respectively. (a) $f = 0.4$; (b) Analytically calculated transmission for $f = 0.4$; (c) $f = 0.5$; (d) $f = 0.6$.

To experimentally investigate the effect of the SWG duty cycle ratio $f$ on the mode-selectivity FoM, two other DUTs with different duty ratios are measured in addition to the one that was discussed above ($f$ = 0.4). The 2D transmission contour map of DUTs with $f$ = {0.5, 0.6} are presented in Fig. 5(c) and Fig. 5(d), respectively. The mode-selectivity FoM is similarly extracted being $\zeta$ = {1.31 ± 0.03, 1.20 ± 0.05}, correspondingly. According to the previous analytical discussions supported by 2D simulation results, the mode-selectivity is expected to be diminished as the duty ratio increases, which is in compliance with the experimentally observed behavior.

In contrast to the validation measurement results of $\zeta$ = {1.44 ± 0.05, 1.31 ± 0.03, 1.20 ± 0.05} for all three DUTs with f = {0.4, 0.5, 0.6}, respectively, the 3D simulation is carried out as described in section II.B, and a mode-selectivity FoM of $\zeta$ = {1.30, 1.28, 1.18} is obtained. We believe that this relatively small discrepancy originates from a superposition of these fabrication process variations: (1) over etching in the fabrication process, which contributes toward decreasing the duty ratio and hence increasing the FoM; (2) the impact of buried air gaps (voids) in the $SiO_2$ cladding between two consecutive corrugations (shown as $L_{Oxide}$ in Fig. 1 (a)) [28], which lowers the effective refractive index for TE1 and eventually gives rise to the FoM; (3) concaved sharp edges in the SWG inner section, which could contribute toward decreasing the FoM depending on the energy distribution of the TE1 mode profile near the edges (sidewall surface roughness could result in spatial field energy redistribution of the TE1 mode); (4) convex sharp edges in the SWG outer section could act otherwise with a very minor effect on FoM expanding due to TE1 energy concentration closer to the core waveguide.

Performance evaluation of the three discussed DUTs in terms of crosstalk and insertion loss require deployment of the mentioned testbed (Fig. 4 with $S_2$ = 10:90), while the 10 % portion being used for optical power loss normalization through a loop-back. In each measurement, only one mode is passing through the DUT for accurate crosstalk monitoring. More specifically, both ports of the DUT through the output mode demultiplexer is monitored, one port being the primary output, and the other as the crosstalk signal. The output transmission is normalized to the spectrum of an on-chip grating coupler loopback interconnecting two back-to-back mode multiplexers to accurately evaluate the DUT insertion loss (IL) at the target wavelength (1540 nm). To remove the impact of Fabry-Perot cavity induced fringes, a moving window of 2 nm is deployed as an averaging filter [11]. The resultant spectrums for the DUTs with different SWG duty ratio $f$ = {0.4, 0.5, 0.6} are illustrated in Fig. 6. Evaluating the IL for the TE1 mode profile shows a distinct IL increase with the SWG duty ratio, which could be explained by the growing TE1 mode redistribution to TE0 (and hence the IL) as the duty ratio enlarges. However, this may not totally explain the relatively larger insertion loss for $f$ = 0.6, which could be ascribed to a fabrication variation induced increased loss of the input(output) mode (de)multiplexer, MMIs, tapers, and S-bends of this DUT. Detailed experimental results are summarized in Table I below including an averaged measured insertion loss over a 40 nm optical bandwidth from 1520 nm to 1560 nm.

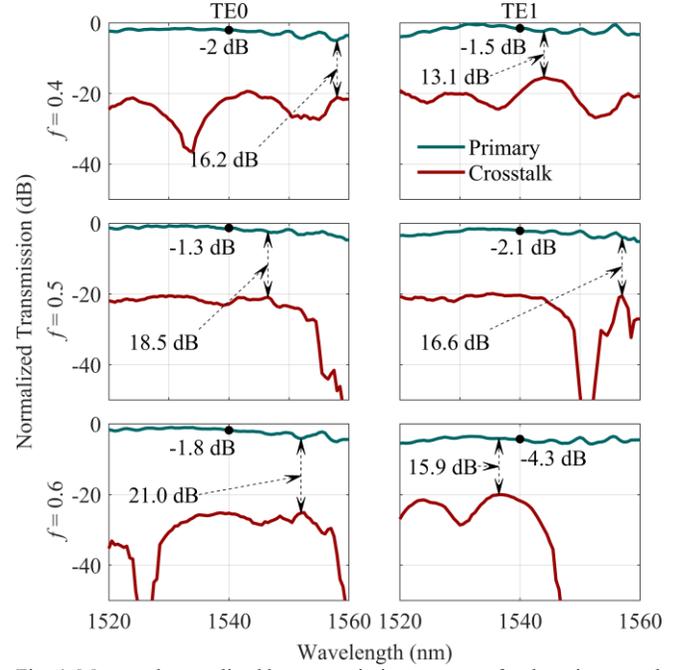

Fig. 4: Measured normalized bar transmission spectrum for the primary and crosstalk mode profiles for DUTs with different SWG duty ratios as $f$ = {0.4, 0.5, 0.6}.

The relatively low amounts of crosstalk and IL in the measured wavelength range (1520 – 1560 nm) provides a promising performance of the DUT for wide-band applications. In addition, because the input(output) mode-(de)multiplexers are the main contributor to the crosstalk, one can expect realization of better crosstalk by utilizing high performance on-chip mode-(de)multiplexers.

## IV.  APPLICATIONS

The proposed MS-TOPS provides a sole degree of freedom to adjust the phase of both TE0 and TE1 modes dependently. Using Eq. (1-2), and noting that the mode-selective TOPS experience almost the same amount of temperature variation $\Delta T$ for both modes, parameter $\zeta$ relates both phase shifts through the relation $\Delta\Phi_{TE0} = \zeta \cdot \Delta\Phi_{TE1}$. Based on the target application, the second degree of freedom can be attained by deployment of an MI-TOPS either in a cascaded configuration, as schematically shown in the subset of Fig. 7, or in an MZI (similar to our proposed test structure) in conjunction with the MS-TOPS.

TABLE I
EXPERIMENTAL (EXP.) AND SIMULATION (3D SIM.) RESULTS FOR THREE DUTs WITH DIFFERENT DUTY CYCLE RATIOS AT THE 1540 NM WAVELENGTH

| Duty ratio ($f$) | Mode-selectivity ($\zeta$) | | Insertion loss (EXP.) (TE0/TE1) [dB] | | Worst-case crosstalk (EXP.) (TE0/TE1) [dB] |
|---|---|---|---|---|---|
| | 3D SIM. | EXP. | @ 1540 nm | Averaged [1520 :1560] nm | |
| 0.4 | 1.30 | 1.44 ± 0.05 | 2/ 1.5 | 2.49/ 1.94 | -16.2/ -13.1 |
| 0.5 | 1.23 | 1.31 ± 0.03 | 1.3/ 2.1 | 1.81/ 2.76 | -18.5/ -16.6 |
| 0.6 | 1.18 | 1.20 ± 0.05 | 1.8/ 4.3 | 2.22/ 4.5 | -21.0/ -15.9 |

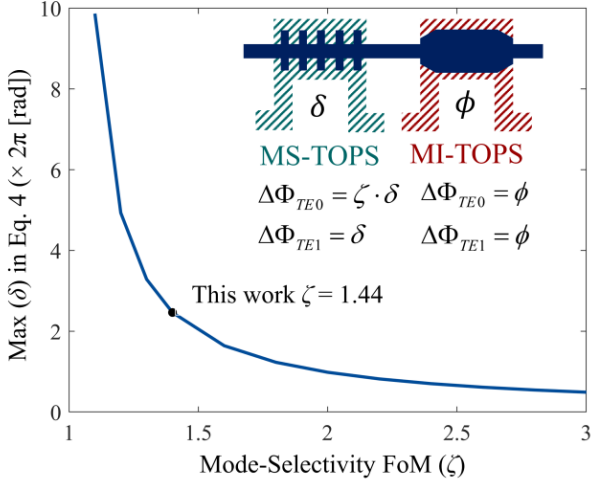

Fig. 5: Driving efficiency analysis of a cascaded MI-TOPS, and MS-TOPS for the application where two degrees of freedom is needed to control both TE0 and TE1 phase shifts with a mode-selectivity of $\zeta$ as shown in the figure subset. Maximum required $\delta$ phase shift quantity is shown versus different mode-selectivity amounts ($\phi \in [0:2\pi]$). In this work ($\zeta$ = 1.44), the MS-TOPS phase shifter needs to be driven up to a maximum of $2.4 \times 2\pi$ to support all phase shift combinations for TE0 and TE1 with two degrees of freedom.

Considering the TE1 phase shifts of $\phi$ and $\delta$ imposed by the cascaded MI-TOPS, and MS-TOPS modules, respectively, the total undertaken phase shifts are formulated as:

$$\Delta\Phi_{total} = \begin{cases} \phi + \zeta\delta, & TE0 \\ \phi + \delta, & TE1 \end{cases} \quad (4)$$

The system of a cascaded mode-selective and insensitive TOPS can realize any desired phase shifts for TE0, and TE1 modes independently. Considering $\phi \in [0:2\pi]$, an arbitrary total phase shift may demand the other phase shifter to be driven further than $2\pi$ for a mode-selectivity FoM of less than two. In other words, an amount of $\zeta = 2$ would facilitate attaining all TE0, and TE1 phase shift combinations while keeping both $\phi$ and $\delta$ within the range of $[0:2\pi]$, which directly improves the system power consumption efficiency. Therefore, the maximum amount of phase shift $\delta$ is a function of the mode-selectivity as illustrated in Fig. 7 as a phase shifter driving efficiency analysis. Considering our work, $\zeta = 1.44$, all TE0, and TE1 phase combinations can be realized with driving the MS-TOPS to a maximum amount of $2.4 \times 2\pi$. A well-optimized mode-selectivity would decrease this amount toward a more power efficient design. Although increasing the mode-selectivity further than $\zeta = 2$ would be optimal from the driving efficiency perspective, depending on the target application, the whole system needs to be engineered to attain the desired phase adjustment precision. The reason being that as the more significant the mode-selectivity is, the more sensitive the system becomes to the applied voltage such that a relatively lower error in the applied voltage translates into a proportionally larger deviation in the targeted phase.

*A. Datacom (switching)*

Considering data communication applications enabled by reconfigurable multimode silicon photonics such as optical networks [29], it may be necessary to route TE0 to a target destination and TE1 to another. A $2\times 2$ MZI incorporating an MS-TOPS in one arm and an MI-TOPS in another could efficiently act as an on-chip mode switch routing TE0 to the first output in bar state ($\pi$ phase shift for TE0 between two arms) and TE1 into the second output in cross state (zero TE1 phase shift for TE1 between two arms). Because it would be possible to simultaneously generate $\pi$ and zero phase shifts between two arms for TE0, and TE1 correspondingly, this space mode switch can separate and route TE0 and TE1 into different paths at the same time, which provides a direct switching methodology of modes without converting them to their fundamental mode.

*B. Scaling up optical processors*

Our research group previously demonstrated a four-by-four reconfigurable MZI-based linear optical processor [30] and experimentally validated implementation of a single layer optical neural network trained for classification of a Gaussian dataset [13]. Scaling-up this topology to higher radix optical processor is challenging due to high insertion loss induced by an increased number of MZI building blocks in each input-output path [31]. Introducing a cascaded configuration of an MS-TOPS and an MI-TOPS could provide the core-enabling building block for realizing a multimode optical processor that can address the scalability bottleneck. With this approach, a radix four multimode optical processor can theoretically realize two simultaneous four-by-four photonic-based mathematical multiplications.

*C. Multimode quantum optical processors*

A recent work has used multi-transverse optical modes for encoding a two-qubit quantum gate [32]; however, it deploys a non-reconfigurable mode-selective directional coupler to solely create a duplication of the signal and attain an additional degree of freedom. In contrast, the proposed MS-TOPS building block can directly manipulate relative phase of TE0 and TE1 modes; which could be cascaded with a MI-TOPS and provide the second degree of freedom. In addition, this could help with efficient realization of transverse mode entangled quantum photonics with supporting the required manipulation of optical modes.

*D. Programming MZI-based optical processors*

Conventional MZI-based optical processors deploy an interconnected mesh of two-by-two MZI building blocks that incorporates an internal and an external phase shifter to adjust output light intensity and phase, respectively. Calibration and accurate programming of the external phase shifters demands an output phase detection technique such as the external coherent detection. We have previously proposed a novel integrated methodology to convert the optical phase to optical power using an MZI incorporating an MS-TOPS and an MI-TOPS in each arm [17]. This enables a low cost, and accurate on-chip optical phase measurement without the need for external coherent detection. Our proposed SWG-based MS-TOPS in this work improves the device mode-selectivity by approximately 32%, which increases the dynamic range and accuracy of the phase measurement task for calibration of multi-transverse mode optical processors.

## V. CONCLUSION

We propose a mode-selective thermo-optic phase shifter (MS-TOPS) exploiting subwavelength grating (SWG) structures. The proposed device can manipulate the relative phase of fundamental transverse electric mode (TE0), while inducing a 1.44 times smaller phase shift in the second mode (TE1) because of the modal field distribution of TE0 in the non-corrugated (center) part of the waveguide and TE1 mode mostly in the SWG section resulting in different thermo-optic coefficients for each mode. This novel device serves numerous applications in mode division multiplexing systems (e.g., optical switching and datacom) by direct mode selective manipulation of the relative phase of each mode instead of using a multitude of components to convert the modes down to their fundamental mode, conduct the desired processing task (e.g., phase shift), and then converting them back to their original mode profile. The balanced MZI test structure incorporating an MS-TOPS, and an MI-TOPS in each arm, and two multimode interferometers demonstrates an insertion loss of 2 (1.5) dB for TE0 (TE1) modes at 1540 nm wavelength. Furthermore, a cascaded configuration of an MS-TOPS and a mode-insensitive thermo-optic phase shifter can adjust the relative phase of each mode independently. A mode-selectivity factor of two can realize all phase combinations while driving each phase shifter within the range of $[0:2\pi]$. Such a device provides the core-enabling building block to scale-up conventional MZI-based optical processors using multimode operation.

ACKNOWLEDGMENT

Authors acknowledge the help from Canadian Microelectronic Corporation (CMC) for the subsidized multi-project wafer fabrication through Applied Nanotools (ANT) as well as financial support from the Natural Sciences and Engineering Research Council of Canada (NSERC).

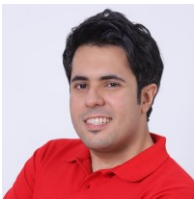

**S. Mohammad Reza Safaee** (M'22) received both his B.S. and M.S. degrees from the Department of Electrical Engineering at Iran University of Science and Technology, Tehran, Iran, in 2013 and 2015, respectively, with a major in electronics. He is currently a Ph.D. candidate at McGill University, affiliated with the Photonics Systems Group within the Department of Electrical and Computer Engineering, located in Montreal, Quebec, Canada. His ongoing research primarily focuses on developing building blocks that enable on-chip optical computing applications, programmable photonics, and photonic digital-to-analog converters.

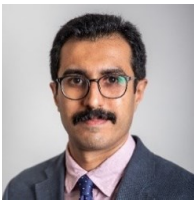

**Kaveh (Hassan) Rahbardar Mojaver** (M'17) received the B.S. and M.S. degrees from Amirkabir University of Technology (Tehran Polytechnic) in 2009 and 2011, respectively, and the Ph.D. degree from Concordia University in 2018, all in electrical engineering. In 2018, he joined the Photonic Systems Group of McGill University as a postdoctoral researcher, where he is currently pursuing his career. His research is focused on photonic integration for optical computing, quantum computing, and data communications.

Dr. Mojaver's research works have been published in several journal and conference publications. He has received the Fonds de Recherche du Québec - Nature et technologies (FRQNT) postdoctoral scholarship, Concordia Merit scholarship, Concordia Accelerator Award, and STARaCom scholarship.

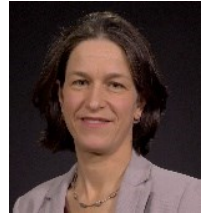

**Odile Liboiron-Ladouceur** (M'95, SM'14) received the B.Eng. degree in electrical engineering from McGill University, Montreal, QC, Canada, in 1999, and the M.S. and Ph.D. degrees in electrical engineering from Columbia University, New York, NY, USA, in 2003 and 2007, respectively. She is currently an Associate Professor with the Department of Electrical and Computer Engineering, McGill University. She was an associate editor of the IEEE Photonics Letter (2009–2016) and was on the IEEE Photonics Society Board of Governance (2016–2018). She holds eight grant U.S. patents and coauthored over 95 peer-reviewed journal papers and 150 papers in conference proceedings. She gave over 30 invited talks on the topic of photonics for computing, photonic computational design methodologies, and photonic-electronic co-design. She is the 2018 recipient of McGill Principal's Prize for Outstanding Emerging Researcher and the 2023 William and Rhea Seath award for engineering innovation.